\documentclass[twocolumn]{aastex62}

\newcommand{\msun}{$M_\sun$}

\newcommand{\kms}{km s$^{-1}$~}

\newcommand{\hi}{\textsc{Hi}}

\defcitealias{janesh15}{J15}
\defcitealias{janesh17}{J17}

\shorttitle{Five Gas-rich UFDs from ALFALFA}
\shortauthors{Janesh et al.}

\begin{document}

\title{Five Gas-rich Ultra-faint Dwarf Galaxy Candidates Discovered in WIYN Imaging of ALFALFA Sources} 

\author{William Janesh} 
  \affiliation{
  Department of Astronomy, Indiana University, 727 E. Third Street,
  Bloomington, IN 47405, USA}
  \affiliation{
  Department of Astronomy, Case Western Reserve University, 10900 Euclid Ave.,
  Cleveland, OH 44106, USA}

\author{Katherine L. Rhode}
  \affiliation{
  Department of Astronomy, Indiana University, 727 E. Third Street,
  Bloomington, IN 47405, USA}

\author{John J. Salzer}
  \affiliation{
  Department of Astronomy, Indiana University, 727 E. Third Street,
  Bloomington, IN 47405, USA}

\author{Steven Janowiecki}
  \affiliation{
  International Centre for Radio Astronomy Research,
  University of Western Australia,
  35 Stirling Highway,
  Crawley, WA 6009, Australia.}
  \affiliation{
  University of Texas at Austin, McDonald Observatory, TX, 79734, USA
  }

\author{Elizabeth A. K. Adams}
  \affiliation{
  ASTRON, The Netherlands Institute for Radio Astronomy, Postbus 2, 7990 AA, Dwingeloo, The Netherlands.}
  \affiliation{
  Kapteyn Astronomical Institute, University of Groningen, Postbus 800, 9700 AV, Groningen, The Netherlands.}

\author{Martha P. Haynes}
  \affiliation{
  Cornell Center for Radiophysics and Space Research, Space Sciences Building, 
  Cornell University, 122 Sciences Drive, Ithaca, NY 14853, USA }

\author{Riccardo Giovanelli}
  \affiliation{
  Cornell Center for Radiophysics and Space Research, Space Sciences Building, 
  Cornell University, 122 Sciences Drive, Ithaca, NY 14853, USA }

\author{John M. Cannon}
  \affiliation{
  Department of Physics and Astronomy, Macalester College, 
  1600 Grand Avenue, Saint Paul, MN 55105, USA }

\begin{abstract}
We present results from the analysis of WIYN pODI imaging of 23 ultra-compact high-velocity clouds (UCHVCs), which were identified in the ALFALFA \hi\ survey as possible dwarf galaxies in or near the Local Group.  To search for a resolved stellar population associated with the \hi\ gas in these objects, we carried out a series of steps designed to identify stellar overdensities in our optical images.  We identify five objects that are likely stellar counterparts to the UCHVCs, at distances of $\sim$350 kpc to $\sim$1.6 Mpc. Two of the counterparts were already described in Janesh et al. (2015) and Janesh et al. (2017); the estimated distance and detection significance for one of them changed in the final analysis of the full pODI data set. At their estimated distances, the detected objects have \hi\ masses from $2\times10^4$$-$$3\times10^6$ \msun, $M_V$ from $-1.4$$-$$-7.1$, and stellar masses from $4\times10^2$$-$$4\times10^5$ \msun. None of the objects shows evidence of a young stellar population.  Their properties would make the UCHVCs some of the most extreme objects in and around the Local Group, comparable to ultra faint dwarf galaxies in their stellar populations, but with significant gas content. Such objects probe the extreme end of the galaxy mass function, and provide a testbed for theories regarding the baryonic feedback processes that impact star formation and galaxy evolution in this low-mass regime.
\end{abstract}


\section{Introduction}
\label{sec:intro}

We have undertaken a campaign to obtain deep, wide-field optical imaging of specific sources detected by the ALFALFA blind neutral hydrogen survey \citep{giovanelli05, haynes11} that are called Ultra-Compact High-Velocity Clouds, or UCHVCs \citep{giovanelli10, adams13}.  These objects have kinematic properties that make them likely to be located in and around the Local Group, but searches of objects in existing optical surveys or catalogs -- in particular, the Sloan Digital Sky Survey \citep[SDSS;][]{sdss}, Digitized Sky Survey (DSS), or the NASA Extragalactic Database (NED) -- reveal no clear optical counterpart. When placed at a fiducial distance of 1 Mpc, UCHVCs have neutral hydrogen properties that are similar to those of gas-rich faint dwarf galaxies like Leo~T \citep{irwin07, ryan-weber08}; specifically, their \hi\ masses are in the $10^5 - 10^6$ \msun\ range, their dynamical masses are $10^7-10^8$ \msun, and their \hi\ diameters are $\sim$2$-$3~kpc \citep{adams13}.  One possibility, explored by \citet{giovanelli10} in the paper that first identified UCHVCs as a category of objects, is that the UCHVCs represent a population of so-called ``minihalos''. These are low-mass ($<10^9$ \msun) dark matter halos that surround detectable baryonic material -- in this case, $\sim10^5-10^6$ \msun\ of neutral gas and perhaps stars as well.

Of the $\sim$30,000 sources detected and cataloged by the ALFALFA survey \citep{giovanelli05, haynes11}, 100 have been identified as UCHVCs to date.  We have been observing a subsample of the UCHVCs with the WIYN 3.5-m telescope\footnote{The WIYN Observatory is a joint facility of the University of Wisconsin– Madison, Indiana University, the University of Missouri, and the National Optical Astronomy Observatory.} in order to search for detectable stellar populations and investigate the exact nature of these objects.  Our subsample consists of the 56 UCHVCs we deem most likely to host a detectable stellar population (see Section \ref{sec:sample} for a discussion of the selection criteria and the detailed properties of these objects).  Identifying a stellar component associated with a given UCHVC in our WIYN imaging allows us to determine a distance to the object, which in turn yields crucial information about the properties (e.g., masses, luminosities) and evolutionary status of these nearby galaxy candidates.  

The first UCHVC we imaged with WIYN led to the discovery of Leo~P, a low-mass, extremely metal-poor dwarf galaxy located at a distance of 1.62$\pm$0.15 Mpc \citep{leop1, leop2, leop3, bc14, mcquinn15}. Finding other such objects is important for a number of reasons. Identifying and studying nearby gas-rich galaxies allows us to investigate the baryonic feedback processes that govern star formation and galaxy evolution in galaxies at the extreme end of the mass function, and it provides important observational data to compare to galaxy formation model predictions \citep[e.g.,][]{revaz18, relhics, vandenbroucke16, verbeke15}.  It also enables us to find some of the lowest-metallicity galaxies in the Universe \citep[e.g.,][]{leop3} and to fill in the low-mass end of the Baryonic Tully-Fisher Relation \citep[BTFR;][]{mcgaugh12}.

This paper is the latest installment in a series of papers presenting the methods and results of our systematic campaign to obtain WIYN imaging of ALFALFA UCHVCs with a wide-field camera first installed at WIYN in 2012 (after the discovery of Leo~P). In the first paper, \citet[][hereafter J15]{janesh15}, we detailed the observational strategy and the procedures we use to search for and characterize the stellar populations associated with the UCHVC targets.  That paper also presented a tentative (92\% significance) detection of a stellar population belonging to the UCHVC AGC~198606. We referred to this object as ``Friend of Leo~T'' because of its proximity in position and radial velocity to the dwarf galaxy Leo~T \citep{irwin07}.  The stellar counterpart we identified for AGC~198606 had a distance of $\sim$380 kpc, a magnitude of $M_V \sim -4.5$, and a large \hi-to-stellar mass ratio \citepalias[between 42 and 110;][]{janesh15}.  The second paper in the series, \citet[][hereafter J17]{janesh17}, described the detection of a stellar population associated with the UCHVC AGC~249525 at a higher significance than that of AGC~198606.  The stellar counterpart in this case had an estimated magnitude $M_V$ between $-4.5$ and $-7.1$ and was more distant, at $1.64 \pm 0.45$ Mpc away.  Its \hi-to-stellar mass ratio was also large, but with a wide possible range (between 9 and 144, depending on how the optical magnitude was estimated).  Though this object has no clearly associated neighbors, the most likely neighbor is the faint irregular galaxy UGC 9128, which has a distance of 2.3 Mpc \citep{tully13} and is $<$10 degrees away in sky position.

Searches by other groups for stellar counterparts to the ALFALFA UCHVCs have yielded varying results.  The SECCO survey \citep{bellazzini15a} observed 25 UCHVCs with the Large Binocular Telescope, detecting a distant ($> 3$ Mpc) counterpart to AGC~226067 that they dubbed SECCO 1, as well as a marginal stellar overdensity associated with AGC~249283. In \citet{adams15a}, our group presented results from \hi\ synthesis observations and WIYN pODI imaging of AGC~226067 and argued that it is associated with the Virgo Cluster rather than being a local object.  \citet{beccari17} used integral field spectroscopy of several \textsc{Hii} regions in SECCO~1; they reiterate that the object indeed likely resides in the Virgo Cluster, and find that it has a much higher metallicity (a mean abundance $\langle$12 + log(O/H)$\rangle = 8.44$) than expected for a dwarf galaxy with a very low mass. 

Using publicly available archival imaging, \citet{sand15} searched for counterparts to objects from both ALFALFA and the GALFA-HI survey \citep{peek11}. \citet{sand15} confirmed the counterpart to AGC~226067 discovered by the SECCO survey, and in a subsequent paper \citep{sand17} confirmed a distance of $\sim$17 Mpc and a relatively high metallicity ([Fe/H] $\sim$ -0.3) based on HST imaging of this object.  Both \citet{beccari17} and \citet{sand17} conclude that AGC~226067 is a pre-enriched remnant gas cloud that was likely stripped from a more massive Virgo Cluster galaxy.  In their search of archival data, \citet{sand15} also identified distant (at least 3 Mpc away) counterparts to four HI-detected objects from the GALFA-HI survey, including two previously identified dwarf galaxies, Pisces A and Pisces B \citep{tollerud15}. HST imaging of the latter two objects yields a tip of the Red Giant Branch (TRGB) distance of 5.6 and 8.9 Mpc, respectively \citep{tollerud16}. The overall result from these other studies is that most of the objects being targeted still have no identified stellar counterpart, and the objects that do have counterparts, while intriguing objects in their own right, are located well outside the Local Group.

In the current paper, we present the entire set of results for the sample of UCHVCs we have imaged with the pODI camera on the WIYN 3.5-m telescope. The pODI camera was mounted on WIYN from 2012 to 2015 and provided a $\sim24\arcmin \times 24\arcmin$ field of view. We observed 23 of the UCHVCs with pODI and now have completed the analysis of the entire pODI data set, finding a total of five dwarf galaxy candidates that we judge to be significant detections and four more candidates that we classify as marginal detections. 

Note that as part of this process, we re-analyzed the data for AGC~198606. In the final analysis AGC~198606 is still classified as a significant detection, although the properties and location of the detection in the AGC 198606 field have changed. 

The remaining objects in our sample of 56 high-priority UCHVCs are being imaged with the ODI camera, which was installed at WIYN in 2015 and provides a $40\arcmin \times 48\arcmin$ contiguous field-of-view.  The analysis of the ODI data is ongoing and the results will be presented in a subsequent paper. 

The paper is organized as follows:  in Section \ref{sec:sample}, we describe the selection criteria and the properties of the sample of ALFALFA UCHVCs we chose for optical follow-up. Section \ref{sec:methods} contains descriptions of our observations and a summary of our data reduction and analysis procedures. We present our results in Section \ref{sec:results} and discuss their implications in Section \ref{sec:discussion}.

\section{The Sample of UCHVCs Observed with WIYN pODI}
\label{sec:sample}
We assembled a sample of UCHVCs for optical follow-up from the catalogs presented in \citet{adams13} and \citet{adams16}. The primary criteria that define whether a source identified in the ALFALFA survey is a UCHVC are given in \citet{adams13}; they are: 
\begin{enumerate}
  \item \hi\ major axis $a <$ 30\arcmin, corresponding to a 2 kpc diameter at 250 kpc, 9 kpc at 1 Mpc, and 22 kpc at 2.5 Mpc. The \hi\ major and minor axes $a$ and $b$ are determined by fitting an ellipse to the \hi\ flux distribution at the level enclosing half of the total flux density \citep[see][]{adams13}.
  \item ALFALFA signal-to-noise ratio $>$ 8, to ensure that the \hi\ signal is a robust detection.
  \item No more than three neighboring \hi\ structures within 3\arcdeg\ on the sky or 15 \kms\ in velocity. A ``most isolated'' subsample is also defined, including sources with no more than four neighbors within 10\arcdeg. UCHVCs must also have no less than 15\arcdeg\ separation from previously known HVC complexes.
  \item Heliocentric radial velocity $-500 < v_{\odot} < 1000$ \kms to select objects within the Local Volume.
  \item Velocity with respect to the Local Standard of Rest, $|v_{lsr}| > 120$ \kms to prevent confusion with Galactic high velocity clouds. Since some similar objects to UCHVCs (e.g., the gas-rich dwarf galaxy Leo~T) have $|v_{lsr}| < 120$ km~s$^{-1}$, this criterion is relaxed \textit{only} if all of the other criteria are met, in order to ensure that we do not miss any legitimate dwarf galaxy candidates.
  \item Critically, \textit{no clear optical counterparts} in SDSS or DSS.
\end{enumerate}

These criteria are intended to select the best possible candidates matching the minihalo hypothesis \citep{giovanelli10}. There are 100 objects that meet these criteria, including the sources with the relaxed velocity criterion (\#5 in the above list) and Leo~P \citep{leop1}. We had access to all of the low-velocity cubes from the ALFALFA survey \citep[as described in][]{haynes18}, and therefore this list of 100 UCHVCs is unlikely to get larger, unless further investigation leads us to relax or otherwise alter the selection criteria. 

\begin{figure*}
  \includegraphics[width=\textwidth]{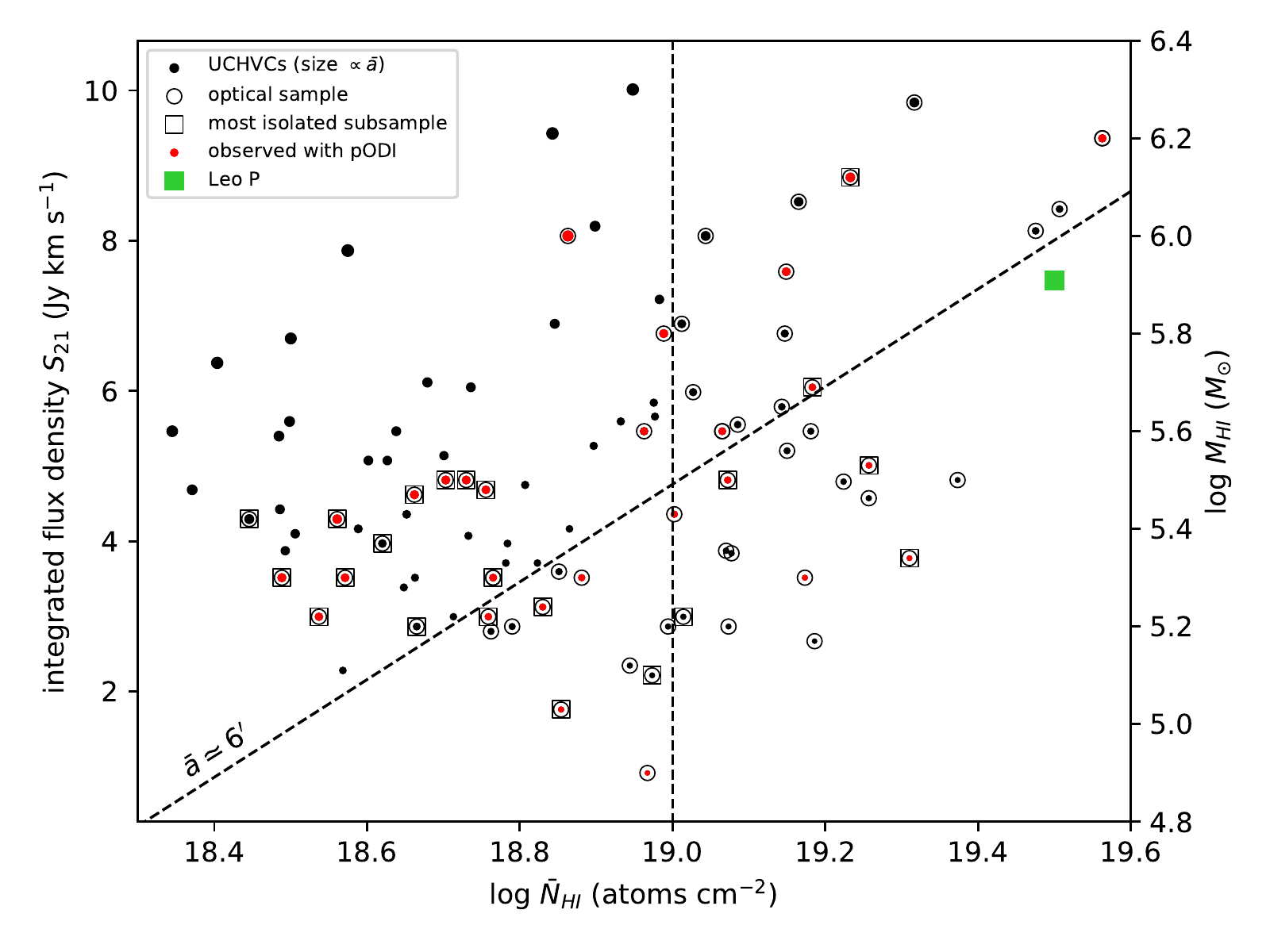}
  \caption{Optical sample selection for UCHVCs using criteria of mean column density ($\bar{N}_{HI}$) and integrated flux density ($S_{21}$), which scales linearly with \hi\ mass at 1 Mpc ($M_{HI}$). All 100 UCHVCs are shown here, with point sizes varying according to their mean angular size as measured in ALFALFA. The dashed lines provide visual references for the criteria in mean column density and mean angular diameter that we use to select the optical sample as described in Section \ref{sec:sample}.  Points outlined in squares belong to the ``most isolated'' subsample. The circled points indicate the objects selected for the optical sample. Red points show the objects we have observed to date with pODI that are presented in this work. Leo P is marked with a large green square at its measured \hi\ mass derived from the distance determination in \citep{mcquinn15}. \label{fig:sample}}
\end{figure*}

In Figure \ref{fig:sample}, we show the full UCHVC sample, illustrated by each object's properties derived from the ALFALFA observations. On the vertical axis, we plot the integrated \hi\ line flux density $S_{21}$ in Jy \kms, and on the horizontal axis we plot the logarithm of the mean column density, log $\bar{N}_{HI}$, in atoms cm$^{-2}$. As an alternate vertical axis we plot the logarithm of the \hi\ mass in solar masses, log $M_{HI}$, at a fiducial distance of 1 Mpc. The size of the points is proportional to $\bar{a} = \sqrt{ab}$, the average angular diameter, and UCHVCs belonging to the most isolated subsample are indicated with a square outline. The \hi\ masses and mean column densities are derived from the equations given in \citet{adams13}:

\begin{equation}
    \bar{N}_{HI} = 4.4 \times 10^{20} \bar{a}^{-2} S_{21} \ \mathrm{cm}^{-2}
\end{equation} 

\begin{equation}
    M_{HI} = 2.356 \times 10^{5} S_{21} d^{2}
\end{equation} 

where $d$ is the measured or assumed distance to the UCHVC.

From the full UCHVC sample shown in Figure \ref{fig:sample}, we selected a subsample of objects to prioritize for deep optical follow-up observations, assuming that the most compact, most isolated, and highest column density objects are more likely to have a stellar counterpart. The objects in this subsample match at least one of the following criteria: 
\begin{enumerate}
  \item Mean \hi\ column density $\log \bar{N}_{HI} > 19$
  \item Mean angular diameter of the half-flux ellipse ($\bar{a} = \sqrt{ab}$) $< 6$\arcmin
  \item Object is in the ``most isolated'' subsample, regardless of its other properties
\end{enumerate}
These criteria are indicated with the vertical and diagonal dashed lines in Figure \ref{fig:sample}. Added to this list were three UCHVCs that do not otherwise meet the criteria above. These three UCHVCs were selected from a master list of ALFALFA UCHVCs provided by members of our collaboration early in the project. They had already been observed with pODI before we defined the specific criteria used to select the highest-priority subsample. All together, the optical follow-up sample contains 59 UCHVC targets, indicated in Figure \ref{fig:sample} with a circle outline. The 23 UCHVCs presented in this work are listed in Table \ref{tab:podidata}. Along with primary and secondary object names, we list the main ALFALFA-derived properties of the UCHVCs: the recessional velocity $cz$, the \hi\ line full-width at half-maximum (FWHM) $w_{50}$, the mean column density $\bar{N}_{HI}$, the mean angular diameter $\bar{a}$, the \hi\ mass at an assumed distance of 1 Mpc, and finally the date the object was observed with pODI. These objects are also marked in red in Figure \ref{fig:sample}; we note that they approximately span the range of the derived properties of the full UCHVC sample.

\begin{deluxetable*}{lllCrrrrrl}
\tabletypesize{\small}
\tablecaption{ALFALFA properties of UCHVCs observed with pODI\label{tab:podidata}}
\tablehead{
  \colhead{Name} & \colhead{R.A.} & \colhead{Dec.} & \colhead{$S_{21}$} & \colhead{$cz$} & \colhead{$w_{50}$} & \colhead{$\bar{a}$} & \colhead{log $\bar{N}_{HI}$} & \colhead{log $M_{HI}$\tablenotemark{a}} & \colhead{Date}\\
  \colhead{} & \colhead{} & \colhead{} & \colhead{[Jy km s$^-1$]} & \colhead{[km s$^-1$]} & \colhead{[km s$^-1$]} & \colhead{[\arcmin]} & \colhead{[atoms cm$^-2$]} & \colhead{[M$_{\sun}$]} & \colhead{Observed}
}
\startdata
AGC174540 & 07h45m59.9s & +14d58m37s &  2.08 \pm 0.21 &   162 & 23 &  7.75 & 19.2 &  5.7 & 2013 Mar \\
AGC198511 & 09h30m13.2s & +24d12m17s &  0.74 \pm 0.07 &   152 & 15 &  6.93 & 18.8 &  5.2 & 2013 Mar \\
AGC198606 & 09h30m05.5s & +16d39m03s &  6.73 \pm 0.67 &    53 & 21 &  9.00 & 19.6 &  6.2 & 2013 Mar \\
AGC198831 & 09h59m11.8s & +18d55m35s &  4.24 \pm 0.42 &    84 & 22 & 16.00 & 18.9 &  6.0 & 2014 Mar \\
AGC208747 & 10h37m06.6s & +20d30m58s &  2.68 \pm 0.27 &    98 & 23 & 11.00 & 19.0 &  5.8 & 2013 Mar \\
AGC208753 & 10h49m32.4s & +23d56m38s &  4.24 \pm 0.42 &    65 & 23 & 13.00 & 19.0 &  6.0 & 2014 Apr \\
AGC215417 & 11h40m08.1s & +15d06m44s &  0.70 \pm 0.07 &   216 & 17 &  9.49 & 18.5 &  5.2 & 2013 Mar \\
AGC219656 & 11h51m24.3s & +20d32m20s &  0.85 \pm 0.08 &   192 & 21 &  8.00 & 18.8 &  5.3 & 2013 Mar \\
AGC226067 & 12h21m54.7s & +13d28m10s &  0.93 \pm 0.09 &  -128 & 54 &  4.47 & 19.3 &  5.3 & 2013 Mar \\
AGC227987 & 12h45m29.8s & +05d20m23s &  5.60 \pm 0.56 &   265 & 26 & 12.00 & 19.2 &  6.1 & 2014 Mar \\
AGC229326 & 12h27m34.7s & +17d38m23s &  0.85 \pm 0.08 &   242 & 23 &  8.00 & 18.8 &  5.3 & 2013 Apr \\
AGC238626 & 13h03m51.1s & +12d12m23s &  0.34 \pm 0.03 &   211 & 36 &  4.00 & 19.0 &  4.9 & 2013 Mar \\
AGC238713 & 13h30m43.8s & +04d13m38s &  1.25 \pm 0.13 &   308 & 26 & 10.95 & 18.7 &  5.5 & 2013 Mar \\
AGC249000 & 14h07m21.7s & +15d45m26s &  1.69 \pm 0.17 &    66 & 31 &  9.00 & 19.0 &  5.6 & 2014 Apr \\
AGC249282 & 14h23m21.2s & +04d34m37s &  1.44 \pm 0.14 &   203 & 40 &  5.92 & 19.3 &  5.5 & 2013 Mar \\
AGC249320 & 14h06m00.0s & +06d07m36s &  0.85 \pm 0.08 &    59 & 24 &  5.00 & 19.2 &  5.3 & 2014 Mar \\
AGC249323 & 14h19m48.6s & +07d11m15s &  1.34 \pm 0.13 &   246 & 20 & 10.82 & 18.7 &  5.5 & 2013 Apr \\
AGC249525 & 14h17m50.1s & +17d32m52s &  6.73 \pm 0.67 &    48 & 24 &  9.00 & 19.6 &  6.2 & 2013 Apr \\
AGC258237 & 15h07m23.0s & +11d32m56s &  1.34 \pm 0.13 &   155 & 23 &  7.07 & 19.1 &  5.5 & 2013 Mar/Apr \\
AGC258242 & 15h10m00.6s & +11d11m27s &  0.70 \pm 0.07 &   207 & 21 &  7.35 & 18.8 &  5.2 & 2013 Mar \\
AGC258459 & 15h04m41.3s & +06d12m59s &  1.34 \pm 0.13 &   149 & 24 & 10.49 & 18.7 &  5.5 & 2014 Mar \\
AGC268069 & 16h05m32.6s & +14d59m20s &  1.14 \pm 0.11 &   132 & 29 &  7.07 & 19.0 &  5.4 & 2013 Mar/Apr \\
AGC268074 & 16h24m43.4s & +12d44m12s &  1.28 \pm 0.13 &   107 & 23 &  9.95 & 18.8 &  5.5 & 2014 Apr \\
\enddata
\tablenotetext{a}{The \hi\ mass at an assumed distance of 1 Mpc}
\end{deluxetable*}

\section{Observations, Data Reduction, \& Analysis}
\label{sec:methods}
Using WIYN and pODI, we observed the targets listed in Table \ref{tab:podidata} between 2013 March and 2014 October, with the bulk of the observations being completed in 2013 March-April. The pODI instrument consisted of nine orthogonal transfer arrays (OTAs) in a $3 \times 3$ configuration that were used for observing targets, and four additional OTAs at larger distances from the field center used for guiding. The central $3 \times 3$ array of OTAs imaged a $\sim24\arcmin \times 24\arcmin$ area on the sky with a pixel scale of 0.11\arcsec\ per pixel. For each target we obtained nine 300s exposures in each of two filters, SDSS $g$ and $i$. The exposures were sequenced in a dither pattern intended to provide uniform coverage across the small gaps between the individual CCD detectors and the larger gaps between the OTAs. 

A detailed discussion of the data reduction and analysis methods used for this project is presented in \citetalias{janesh15}, but is summarized here for clarity. The raw images are processed using the QuickReduce pipeline \citep{kotulla14} to remove the instrumental signature, including corrections for bias, flat-fielding, dark current, persistency, fringing in the $i$ images, and the WIYN pupil ghost. To create a final combined image, the QuickReduce-processed images are illumination-corrected \citep[see][]{janowiecki15}, reprojected to a common pixel scale, scaled to a common flux level, and stacked. The combined images are then aligned and trimmed to a common size of 11,000 pixels by 11,000 pixels ($\sim20\arcmin \times 20\arcmin$) to remove the high-noise regions on the edges of the dither pattern. 

For each pair of $g$ and $i$ images, we then construct a stellar catalog. First, sources in each image are detected using \texttt{daofind} in IRAF\footnote{IRAF is distributed by the National Optical Astronomy Observatory, which is operated by the Association of Universities for Research in Astronomy (AURA) under a cooperative agreement with the National Science Foundation.}, with a threshold level typically 3.5 times the standard deviation of the background counts in each image. Then sources are matched in the two filters to eliminate spurious detections. 

We eliminate extended sources by first measuring the instrumental magnitude within an aperture with radius equal to the average FWHM PSF of the image and another aperture set to twice that value.  For unresolved sources of a given instrumental magnitude, the difference between the measured magnitude within these two apertures ($m_{2\times}-m_{1\times}$) will be approximately constant. 
We compute these values for each set of images, then in a series of instrumental magnitude bins, select only the sources within three standard deviations of the relevant value. The extended source cut described here differs from that described in \citetalias{janesh15}, and allows us to apply a consistent and predictable correction using the statistical method rather than a by-eye selection of point sources. By using the characteristics of known bright point sources, we can more easily reject extended sources at fainter magnitudes. The data presented in \citetalias{janesh15} have been reanalyzed in this work for consistency with the rest of the optical follow-up sample.

We carry out photometry of all the remaining point sources with a small aperture, apply an aperture correction, and calculate calibrated $g$ and $i$ magnitudes by determining calibration coefficients (zero points and color terms) based on SDSS stars in each image and applying them to the instrumental magnitudes we measure. Finally, we apply a Galactic extinction correction to the final photometry of the stellar sources. The Galactic extinction correction is calculated by applying the \citet{sf11} coefficients to the \citet{schlegel98} values for the center of the field since the UCHVCs are at high Galactic latitudes and exhibit small variations in reddening across the field.  

We also perform a series of artificial star experiments -- adding artificial point sources with a range of magnitudes to each image and then executing the same detection and photometry steps -- in order to quantify the detection completeness as a function of magnitude for the final stacked $g$ and $i$ images of each object. The detection completeness is indicated in each of the color-magnitude diagrams (CMDs) shown in the figures in Section \ref{sec:results}, as described in the captions.

With a catalog of stellar sources in the field in hand, we then search for a resolved stellar counterpart using a CMD filter, following a similar strategy to that of \citet{wwj09}. The CMD filter is constructed using isochrones from \citet{girardi}, selected to represent old (8-14 Gyr), metal-poor ($Z = 0.0001-0.0004$) populations. Stars are selected by the CMD filter in color-magnitude space at a range of distance moduli, then their positions are spatially binned into a two-dimensional (2D) histogram with 8\arcsec\ $\times$ 8\arcsec\ bins. 
The 2D histogram grid is then smoothed using a Gaussian kernel with a 2\arcmin\ -- 3\arcmin\ FWHM, approximately the range of expected angular sizes for dwarf galaxies over the range of distance moduli. The 2D histogram is then normalized so that the final values in the grid are expressed in standard deviations from the mean; we use this information to identify stellar overdensities. 

We quantify the significance, $\xi$, of the overdensities we find by performing a Monte-Carlo style test, generating 25,000 random, uniformly spatially distributed sets of points, with the number of points set equal to the number of stars that are selected by the CMD filter. Applying the same smoothing and normalization technique to the random datasets allows us to construct a probability distribution of their ``most significant'' overdensities, to which we can compare the overdensities selected in the real data. We express $\xi$ in terms of the cumulative distribution function -- e.g., a $\xi$ of 90\% means that 90\% of the ``most significant'' overdensities in the random datasets are as many or fewer standard deviations from the mean than the overdensity from the real data. Overdensities with a high significance ($\xi>97\%$) are unlikely to be due to random superpositions of stars and are instead likely to be true collections of stars that are physically associated. We also find both the minimum and maximum distance that an overdensity with $\xi > 90\%$ is detected at the same sky position. We then assume the distance error to be the greater of the differences between the distance of the highest significance overdensity and the minimum and maximum distances.

To classify the overdensities, we consider other information in addition to their significance. The position of the overdensity relative to the \hi\ detection is of particular interest. We compute an angular distance from the center of the overdensity to the ALFALFA \hi\ centroid, $\Delta_{HI}$, as a first-order indicator of the coincidence, and also visually inspect the position relative to the ALFALFA \hi\ column density contours, and when available, contours from higher-resolution \hi\ synthesis imaging \citep[e.g., the {\hi} synthesis observations of a subset of UCHVCs from the Westerbork Synthesis Radio Telescope presented in ][]{adams16}. It is reasonable to expect that the detected stellar overdensities that are spatially coincident with the \hi\ gas distribution are more likely to be stellar populations genuinely associated with the UCHVCs. We also consider the proximity of the overdensities and UCHVCs to nearby galaxies and galaxy groups with available literature values for recessional velocity and distance. Detecting a stellar population at a distance similar to a nearby galaxy or group that has a recessional velocity similar to the UCHVC is another indicator that the stellar population is a dwarf galaxy actually associated with that group or galaxy. 

The robustness of a detection is also a factor in its classification. We consider four individual overdensities when determining whether a detection is robust: the highest significance overdensity with a smoothing cell size of 2\arcmin, the highest significance overdensity with a smoothing cell size of 3\arcmin, and the highest significance close overdensity ($\Delta_{HI} < 8.0\arcmin$) at 2\arcmin\ and 3\arcmin\ smoothing cell sizes. We select the highest significance overdensity of these as the primary detection, then use the remaining three as supporting detections. To be a robust overdensity, the primary and at least two of the supporting detections must be located at the same spatial position or within fewer than five bins (40\arcsec) in the smoothed 2D histogram, have individual significances greater than 90\%, and be located within the set of overdensities used to determine the distance error for the primary detection. Detections that are more than 8\arcmin\ from the \hi\ centroid are considered non-robust.

When we do detect an overdensity that we think is a stellar population genuinely associated with a UCHVC, we then estimate the optical and \hi\ properties of the UCHVC based on the distance at which the overdensity was detected. The gas properties -- \hi\ mass, dynamical mass, and physical size -- are directly proportional to the distance, but estimating the optical properties requires further analysis of the WIYN imaging data. We obtain two estimates of the apparent magnitude of the stars in the overdensity. 
The first estimate, which is essentially a minimum brightness, is derived by summing the flux from the individual stars that lie within the 2\arcmin\ smoothing cell centered on the peak of the overdensity.  The second estimate, which amounts to a maximum brightness, is derived by performing photometry within a large (2\arcmin) aperture centered on the overdensity, after masking out any sources that are obviously not part of a potential dwarf galaxy counterpart (e.g., resolved background galaxies, bright foreground stars). The uncertainties on these measurements are based solely on the magnitude errors from the photometry of either the individual stars or the large apertures themselves.
We combine these minimum and maximum brightness estimates with the distance to the object to estimate the possible absolute magnitude range, including calculating the $M_V$ value in order to facilitate comparison to literature values. With an estimate of the luminosity and color of the object, we can determine its mass-to-light ratio using the relations given in \citet{bell03}. The mass-to-light ratio can then be used to compute a stellar mass and an HI-to-stellar mass ratio.

\section{Results}
\label{sec:results}

\begin{deluxetable*}{lrrrrrCRRRRrr}
  \tablewidth{0pt}
  \tablecolumns{13}
\tabletypesize{\small}
\tablecaption{Five Gas-rich, Ultra-faint Dwarf Galaxy Candidates from WIYN pODI Imaging of ALFALFA UCHVCs\label{tab:optical}}
\tablehead{
  \colhead{Name} & \colhead{$d$} & \colhead{$\delta d$} & \colhead{$\xi$} & \colhead{$\Delta_{HI}$} & \colhead{$\bar{a}/2$} & \colhead{log $M_{HI}$} & \multicolumn{2}{c}{$M_V$\tablenotemark{a}}  & \multicolumn{2}{c}{log $M_{*}$\tablenotemark{a}} & \multicolumn{2}{c}{$M_{HI}/M_{*}$\tablenotemark{a}} \\
  \colhead{} & \colhead{} & \colhead{} & \colhead{} & \colhead{} & \colhead{} & \colhead{} & \colhead{(faint)\tablenotemark{a}} & \colhead{(bright)} & \colhead{(faint)} & \colhead{(bright)} & \colhead{(bright)} & \colhead{(faint)}\\
  \colhead{-} & \colhead{[kpc]} & \colhead{[kpc]} & \colhead{[\%]} & \colhead{[kpc]} & \colhead{[kpc]} & \colhead{[M$_{\sun}$]} & \colhead{[mag]} & \colhead{[mag]} & \colhead{[M$_{\sun}$]} & \colhead{[M$_{\sun}$]} & \colhead{-} & \colhead{-}
}
\startdata 
AGC198606                  &  880 &   20 &  97.9 &  1.4 &  1.2 & 6.1 \pm 0.1 & -4.0 \pm 0.01 & -6.7 \pm 0.13 & 3.8 \pm 0.01 & 5.5 \pm 0.10 &  3.8 & 205.0 \\ 
AGC215417                  &  350 &   20 &  98.2 &  0.3 &  0.5 & 4.3 \pm 0.1 & -1.4 \pm 0.03 & -5.2 \pm 0.10 & 2.6 \pm 0.02 & 4.0 \pm 0.04 &  5.3 & 145.5 \\ 
AGC219656                  &  900 &  200 &  99.9 &  0.7 &  1.0 & 5.2 \pm 0.2 & -3.9 \pm 0.02 & -7.0 \pm 0.09 & 3.6 \pm 0.01 & 4.7 \pm 0.02 &  3.3 &  36.9 \\ 
AGC249525\tablenotemark{b} & 1600 &  450 &  98.4 &  0.4 &  2.1 & 6.5 \pm 0.3 & -4.5 \pm 0.02 & -7.1 \pm 0.07 & 4.3 \pm 0.10 & 5.6 \pm 0.06 &  9.0 & 144.0 \\ 
AGC268069                  &  700 &  100 &  99.9 &  0.9 &  0.7 & 5.1 \pm 0.1 & -2.8 \pm 0.03 & -6.9 \pm 0.08 & 3.4 \pm 0.02 & 5.2 \pm 0.05 &  0.8 &  48.8 \\ 
\enddata
\tablenotetext{a}{See Section \ref{sec:methods} for description for methods of determining limits on optical properties.}
\tablenotetext{b}{Values from \citetalias{janesh17}}
\tablecomments{This table lists the highest significance detection at \textit{any} distance in the pODI image for each UCHVC. As described in Section \ref{sec:methods}, these objects are robust overdensities with $\xi > 97\%$ and $\Delta_{HI} < 8.0\arcmin$.} 
\end{deluxetable*}

\begin{deluxetable*}{lrrrrrc}
  \tablewidth{0pt}
  \tablecolumns{7}
\tabletypesize{\small}
\tablecaption{Results of CMD filtering procedure for remaining UCHVC sample observed with WIYN pODI\label{tab:optical2}}
\tablehead{
  \colhead{Name} & \colhead{$d$} & \colhead{$d_{err}$} & \colhead{$\xi$} & \colhead{$\Delta_{HI}$} & \colhead{$\bar{a}/2$} & \colhead{log $M_{HI}$} \\
  \colhead{-} & \colhead{[kpc]} & \colhead{[kpc]} & \colhead{[\%]} & \colhead{[kpc]} & \colhead{[kpc]} & \colhead{[M$_{\sun}$]}
}
\startdata %
\sidehead{Non-Robust Overdensities with $90\% < \xi < 97\%$ and $\Delta_{HI} < 8.0\arcmin$}
\tableline
AGC198831 &  410 &   20 &  92.0 & 0.2 & 1.0 & 5.2  \\
AGC249320 & 1140 &   70 &  95.2 & 2.6 & 0.8 & 5.4  \\
AGC258242 & 1020 &  100 &  90.3 & 2.3 & 1.1 & 5.2  \\
AGC268074 &  260 &  100 &  90.5 & 0.2 & 0.4 & 4.3  \\
\sidehead{Corner and Edge Overdensities ($\Delta_{HI} > 8.0\arcmin$)}
\tableline
AGC174540 & 1100 &  250 &  99.1 & 2.8 & 1.2 & 5.8  \\
AGC198511 & 2230 &  300 &  99.0 & 8.7 & 2.3 & 5.9  \\ 
AGC208747 &  520 &  100 &  99.9 & 2.1 & 0.8 & 5.2  \\ 
AGC208753 &  270 &   40 &  99.8 & 0.8 & 0.5 & 4.9  \\
AGC226067 &  880 &   60 &  98.3 & 2.2 & 0.6 & 5.2  \\
AGC227987 &  820 &  250 &  99.9 & 2.8 & 1.4 & 5.9  \\
AGC229326 &  250 &   20 &  99.7 & 1.0 & 0.3 & 4.1  \\
AGC238626 &  450 &   20 &  99.6 & 1.7 & 0.3 & 4.2  \\
AGC238713 & 1740 &  200 &  97.8 & 4.7 & 2.8 & 6.0  \\
AGC249000 & 1900 &  700 &  99.0 & 6.1 & 2.5 & 6.2  \\
AGC249282 &  630 &  160 &  99.5 & 1.7 & 0.5 & 5.1  \\
AGC249323 & 1900 &  100 &  97.2 & 4.8 & 3.0 & 6.1  \\
AGC258237 & 2400 &  200 &  96.3 & 6.0 & 2.5 & 6.3  \\
AGC258459 & 1340 &  200 &  97.3 & 3.7 & 2.0 & 5.8  \\
\enddata
\end{deluxetable*}

The results of our analysis of the sample of UCHVCs observed with WIYN pODI are shown in Figures \ref{fig:results1}, \ref{fig:results2}, \ref{fig:nondets1}, and \ref{fig:nondets2}, and summarized in Tables \ref{tab:optical} and \ref{tab:optical2}. For each UCHVC, we present here the highest significance overdensity detected. In the Figures, we show the CMD filtering process at the given distance and the smoothed stellar density map.

Each overdensity detected is assigned a category depending on its significance and location in the image relative to the \hi\ distribution. Table \ref{tab:optical} includes robust detections ($\xi > 97\%$) within 8\arcmin\ of the \hi\ centroid. We report the distance $d$ and its error $\delta d$, measured as the range of distances at which an overdensity with a significance $\xi > 90\%$ occurs at the same location; the significance $\xi$; the distance of the overdensity from the ALFALFA \hi\ centroid $\Delta_{HI}$, converted to kpc at the distance of the detection; the mean half-flux radius from ALFALFA $\bar{a}$, converted to kpc at the distance of the detection; the logarithm of the \hi\ mass at the measured distance log $M_{HI}$; and the derived optical properties as described in Section \ref{sec:methods}.

Table \ref{tab:optical2} includes non-robust detections ($\xi$ between 90\% and 97\%) within 8\arcmin\ of the \hi\ centroid, and corner or edge detections, which may be highly significant but are more than 8\arcmin\ away from the \hi\ centroid and are thus on the edge of the pODI image. The 8\arcmin\ radius also corresponds to the largest-sized UCHVC in the sample presented in this paper. We argue later in this section that it is unlikely that a detection of an overdensity that is located several arcminutes away, in the far corner or edge of the image, is truly associated with the UCHVC target identified by ALFALFA. In Table \ref{tab:optical2}, we report the same values as in Table \ref{tab:optical}, but do not estimate the optical properties of these objects. We discuss our results for each of these categories below, giving more detail on individual targets when appropriate.


\subsection{Overdensities with $\xi > 97\%$}

\begin{figure*}
  \includegraphics[width=\textwidth]{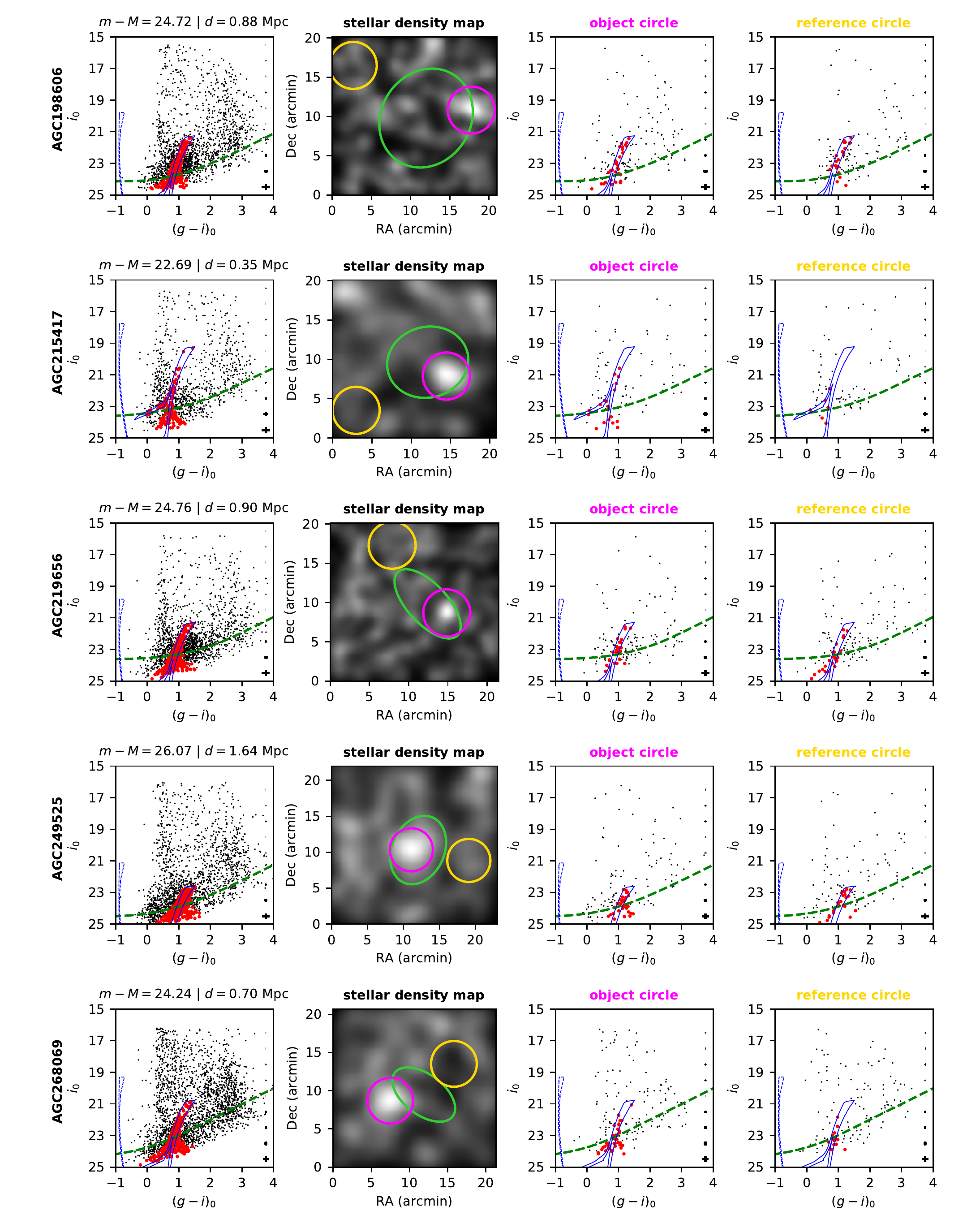}
  \caption{Results from the CMD filtering procedure for the candidate dwarf galaxies AGC~198606, AGC~215417, AGC~219656, AGC~249525, and AGC~268069.  The CMD panels (far left) show the CMD locations of all stars in the field as small black points, with the CMD filter shown as a solid blue line. Stars selected by the CMD filter are shown as larger red points. We show the median instrumental magnitude error in a series of 1 magnitude bins along the right side of each CMD panel. The 50\% completeness level in the $i$-band as a function of $g-i$ color, derived from artificial star tests, is shown with a dashed green line. A filter representative of a young stellar population ($\sim10$ Myr) is shown as a blue dashed line, but is \textit{not} used to select stars. The construction of the CMD filters is described in \citetalias{janesh15}. The smoothed density of stars selected by the CMD filter is shown in the middle left panels. Grayscale color is mapped to the number of standard deviations from the mean density in the field. The magenta circle is centered on the highest density bin in the density map, while the yellow circle is randomly placed as a reference field. The green ellipse is the 21-cm line half-power ellipse \citep[see][]{adams13}. The remaining panels show CMDs (as described in the far left panel) for the regions enclosed by the colored circles: the magenta object circle (middle right) and the yellow reference field circle (far right). \label{fig:results1}}
\end{figure*}

Five objects have $\xi$ values in the range 97.9-99.9\%, distances between 350 kpc and 1.6 Mpc, and are located anywhere from $\sim$1.0\arcmin\ to $\sim$5.6\arcmin\ from the centroid of the \hi\ emission, and often within the \hi\ distribution as characterized by the half-flux ellipse (see Figure \ref{fig:results1}). They have stellar masses between $10^2$ and $10^6$ \msun, \hi-to-stellar-mass ratios between 0.8 and 145, and absolute $V$ magnitudes between -1.4 and -7.1. The estimated magnitude and stellar mass values are consistent with the range found in nearby UFDs, but the $M_{HI}/M_{*}$ values are large, whereas UFDs are typically gas-poor. Unlike Leo~P, these objects appear to contain very few blue stars or stars nearby to the young stellar population isochrones, typically only as many as three or four, when considering the CMDs shown in Figure \ref{fig:results1}, indicating that they have not undergone recent star formation.

One of the high-significance detections listed in Table \ref{tab:optical} is associated with AGC~249525. Results for this object were presented previously in \citetalias{janesh17} and are included in the results presented here. This overdensity is directly coincident with the highest density \hi\ contour derived from Westerbork Synthesis Radio Telescope (WSRT) synthesis imaging, and additionally shows evidence for an overdensity of RGB stars in the center of the image,  without using a CMD filter to select a population at a particular distance. 

\begin{figure*}
  \includegraphics[width=\textwidth]{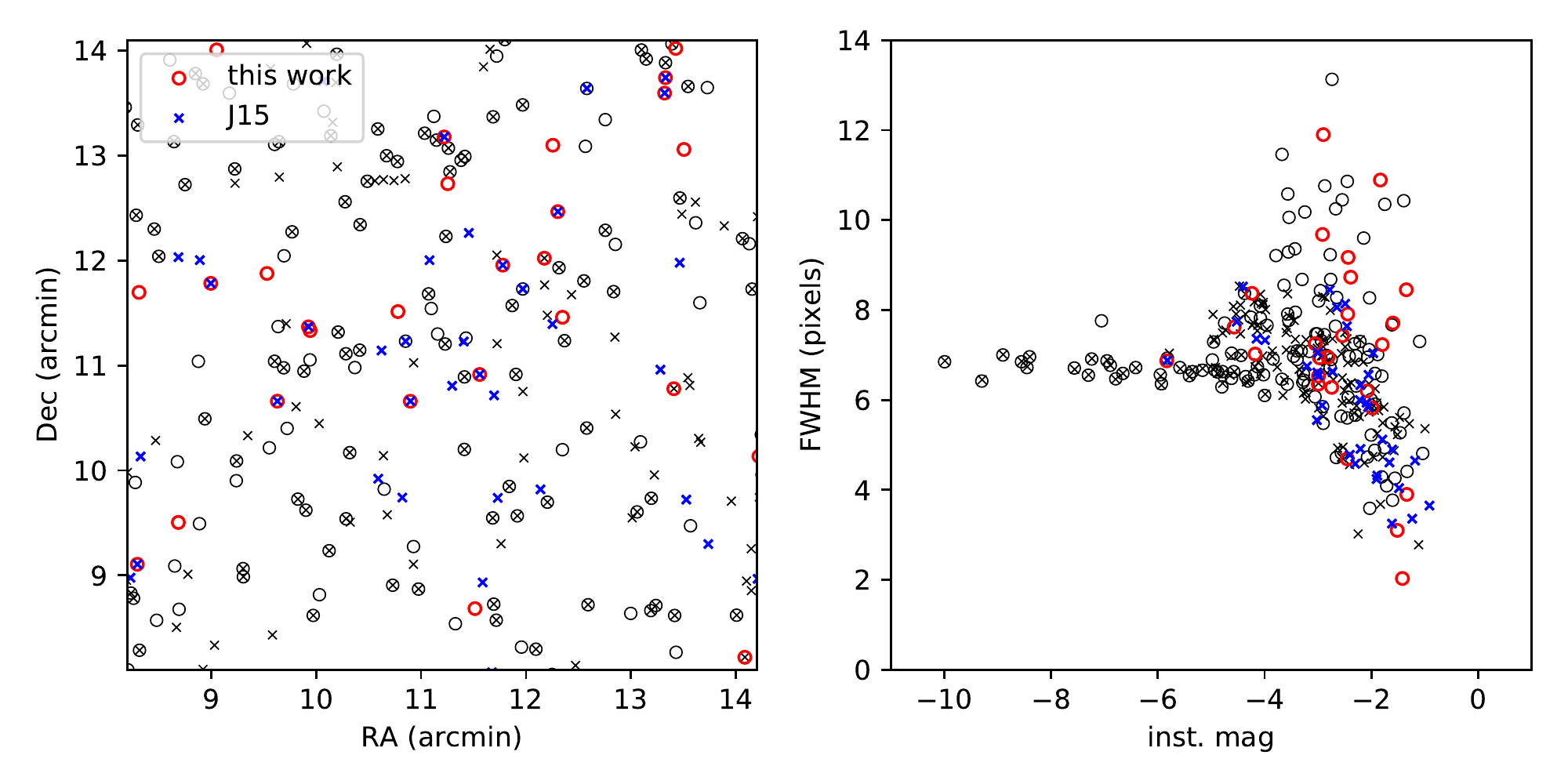}
  \caption{Comparison of extended source cut results for AGC~198606 for this work and \citetalias{janesh15}. The left panel shows the sky positions of sources passing the extended source cut in both works, centered on the location of the overdensity reported in \citetalias{janesh15}. The \citetalias{janesh15} sources indicated with crosses, and the analysis presented in this work with circles. Crosses are highlighted in blue (J15), and circles are highlighted in red (this work) if they are selected by the CMD filter. The \citetalias{janesh15} overdensity can be identified as an increased number of blue crosses in the center of the region shown in the figure. The right panel shows the results of the extended source cut in a plot of FWHM vs. instrumental magnitude but only for sources in the field-of-view of the left panel. The symbols have the same meaning as in the left panel. \label{fig:escut}}
\end{figure*}

As explained in Section \ref{sec:intro}, we also previously reported the detection of a stellar population associated with AGC~198606 in \citetalias{janesh15}, and nicknamed this object ``Friend of Leo T''. At the time, our search algorithm detected the highest significance overdensity ($\xi = 92\%$) at $\sim380$ kpc. In this work, we have applied a different algorithm to remove extended sources from our photometric catalog and reanalyzed the data (see the description of this method in Section \ref{sec:methods}). Figure \ref{fig:escut} compares the \citetalias{janesh15} results with the reanalysis in this work. The left panel shows the stars in the field, with highlighted CMD filter-selected stars composing the overdensity described in \citetalias{janesh15}, as well as stars in the same region from the reanalyzed data selected using an identical CMD filter at the same distance. Stars in common between both \citetalias{janesh15} and this work are identifiable by overlapping symbols. Nearly all of the stars composing the \citetalias{janesh15} overdensity are not present in the reanalysis, having been removed by the revised extended source cut algorithm. The right panel shows the difference in the results of the extended source cut for this portion of the pODI images, showing all the point sources with the stars selected by the CMD filter highlighted, in the $6\arcmin \times 6\arcmin$ region from the left panel. The extended source cut used in the reanalysis allows more faint sources with higher FWHM values, while the \citetalias{janesh15} extended source cut allows for more faint sources with lower FWHM values.

Using the different extended source cut algorithm, we identify a different overdensity as the most significant in the field, located 9\arcmin\ to the West of the previously reported overdensity at a distance of 880 kpc. At a significance of $\xi = 97.9\%$ this overdensity is still potentially associated with the \hi\ distribution, overlapping the half-flux ellipse. This stellar overdensity is now approximately twice the distance of Leo~T, though our argument in \citetalias{janesh15} that AGC~198606 could be associated with Leo~T based on the kinematics of the gas still applies, with the line of sight separation between the gas distributions remaining the same.

AGC~215417 is located at 350 kpc with a significance of $\xi = 98.2\%$. At this distance it is the lowest mass object among the robust detections and is also extremely faint. It is also unlikely to be associated with any of the known galaxies with similar radial velocities ($cz = 215$ \kms) within 10\arcdeg\ on the sky since they are all located at much greater distances (at least 1 Mpc). AGC~215417 was observed in \citet{bellazzini15a} as their object ``L''. Though their analysis did not consider it to be an optical counterpart, their smoothing algorithm detected a region of higher point source density at approximately the same sky position as our overdensity.

AGC~268069 shares the highest significance in our sample ($\xi = 99.9\%$), at a distance of 700 kpc. There are no cataloged galaxy neighbors within 10 degrees that have similar systemic radial velocities \citep[$cz=131$ \kms;][]{adams13} or distances; however, ten other ALFALFA sources with radial velocities between $cz=100$ \kms and $cz=160$ \kms are located less than 6\arcdeg\ away on the sky. AGC~268069's upper limit estimated stellar mass is the second highest of the UCHVCs in Table \ref{tab:optical}, at $1.6 \times 10^5$ \msun. AGC~268069 was observed in \citet{bellazzini15a} as their object ``W'', and was not associated with an optical counterpart in that work. If the imaging data from \citet{bellazzini15a} is centered on the ALFALFA \hi\ centroid of AGC~268069, the overdensity we have identified here is outside of the footprint of those images.

AGC~219656 is the other highest significance detection in the sample ($\xi = 99.9\%$), and is located at a distance of 900 kpc. It has a recessional velocity of 192 \kms, and no obvious neighbors within 10\arcdeg\ of its position that also have a recessional velocity $cz < 500$ \kms. AGC~219656 is included in the sample of UCHVCs observed with WSRT in \citet{adams16}, but was not selected as a good candidate for follow-up based on the higher-resolution WSRT data.

\subsection{Overdensities with $90\% < \xi < 97\%$}

\begin{figure*}
  \includegraphics[width=\textwidth]{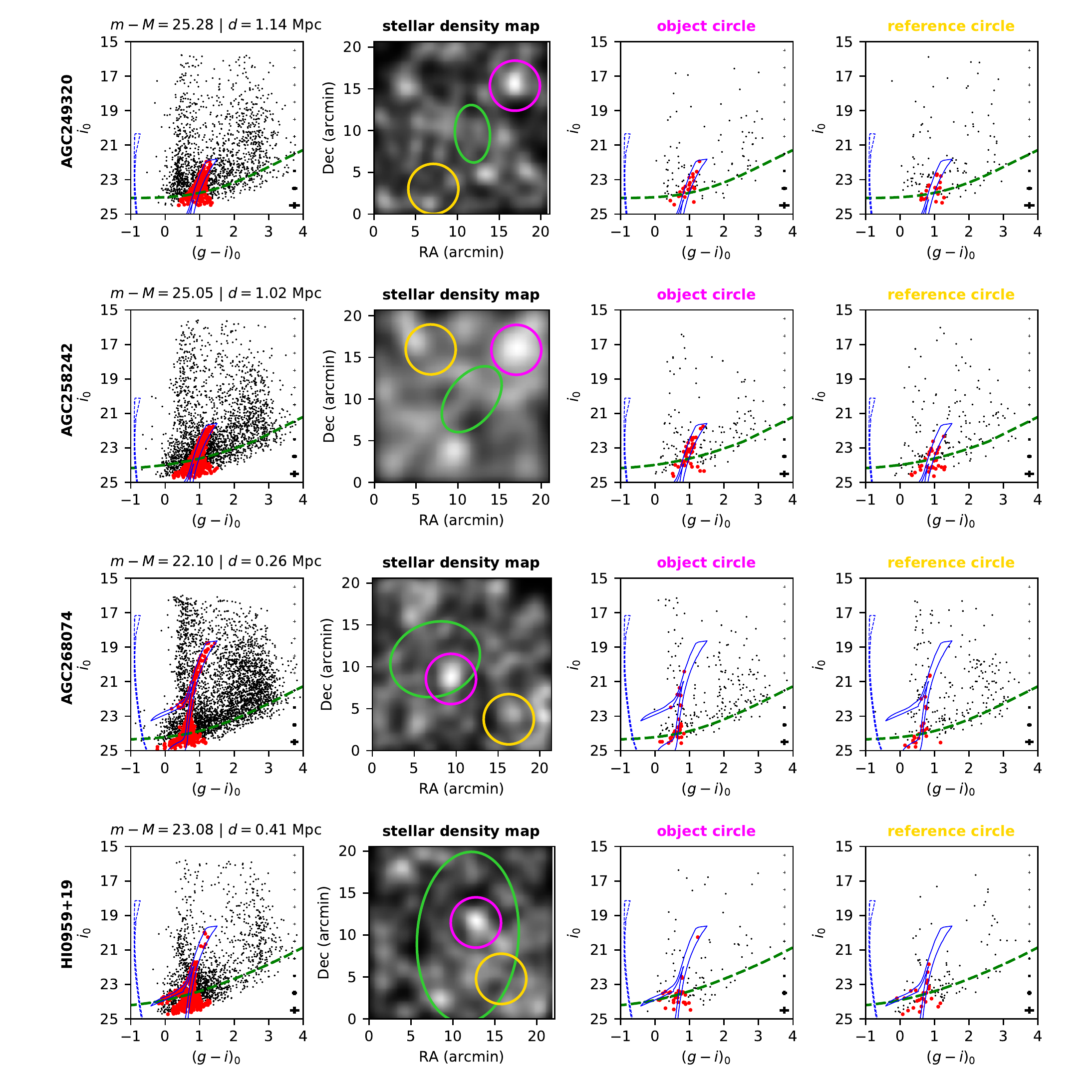}
  \caption{Results from CMD filtering procedure for the non-robust detections AGC~198831, AGC~249320, AGC~258242, and AGC~268074. The panels are the same as described in the caption for Figure \ref{fig:results1}. \label{fig:results2}}
\end{figure*}

Four of the objects are classified in Table \ref{tab:optical2} as non-robust close overdensities and shown in Figure \ref{fig:results2}. These objects all have relatively low significances between $\xi=90.3\%$ and $\xi=95.2\%$ but are located within 8\arcmin\ of the center of the \hi\ distribution. While they do not meet the criteria for robust detections because of their lack of supporting overdensities, these marginal detections are still potential stellar counterparts.

Two of these objects were observed by \citet{bellazzini15a}, AGC~258242 and AGC~268074. Neither of these objects had significant stellar overdensities in the Large Binocular Telescope images from that work, though the footprint of those images is smaller ($14\arcmin \times 7\arcmin$) than that of the pODI images. Our detection for AGC~258242 would be well outside that footprint, assuming the LBT image is centered on the \hi\ centroid. In the case of AGC~268074, the object detected in our pODI images \textit{would} likely fall within the LBT image footprint, but \citet{bellazzini15a} do not detect an overdensity in their data for this UCHVC.

\subsection{Corner and Edge Overdensities}\label{sec:nondets}
\begin{figure*}
  \includegraphics[width=\textwidth]{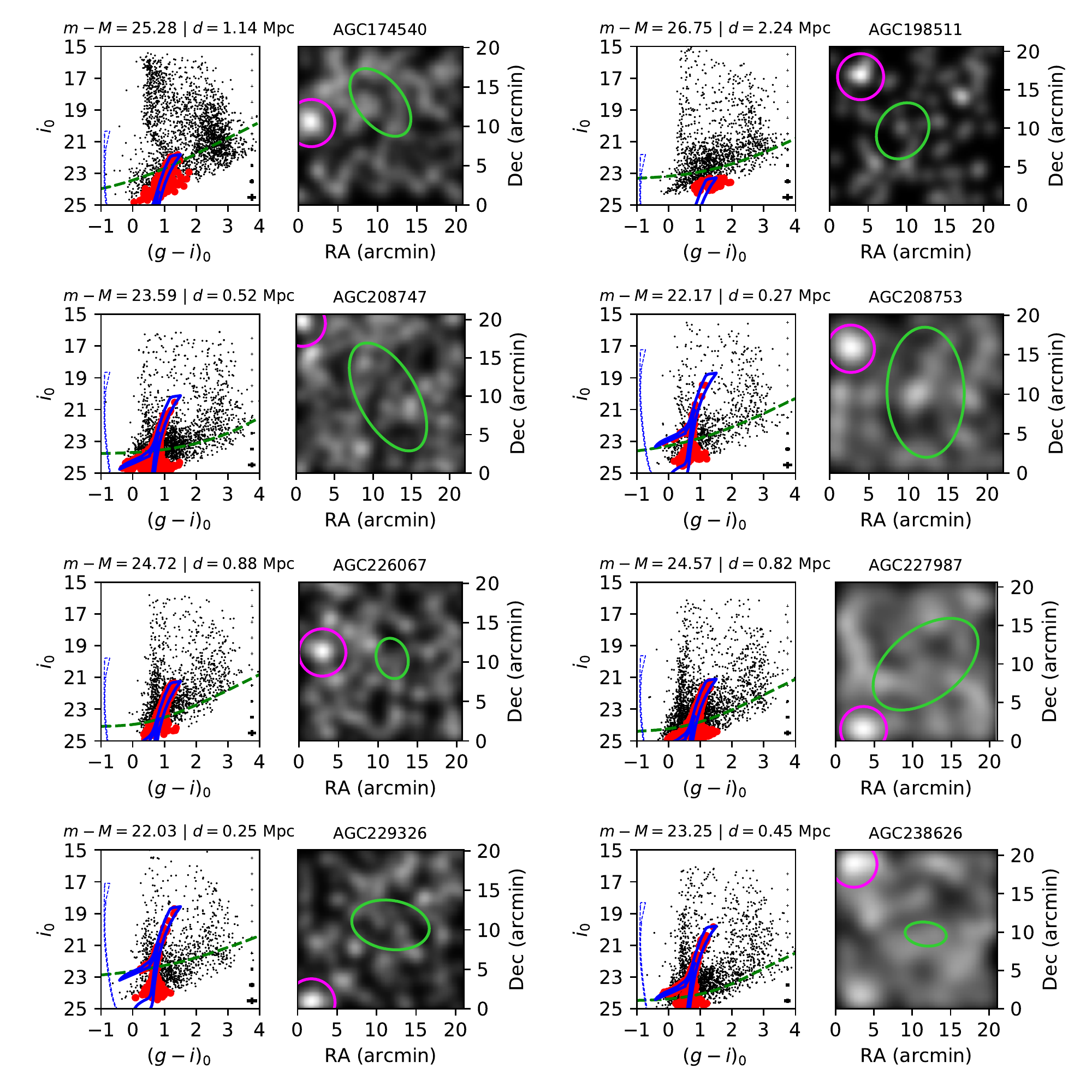}
  \caption{CMDs and smoothed stellar density maps, as described in Figure \ref{fig:results1}, for the first eight of the fourteen objects that only had overdensities detected at the corner or edge of the image, several arc minutes away from the location of the ALFALFA UCHVC. See discussion of these objects in Section \ref{sec:nondets}. \label{fig:nondets1} }
\end{figure*}

\begin{figure*}
  \includegraphics[width=\textwidth]{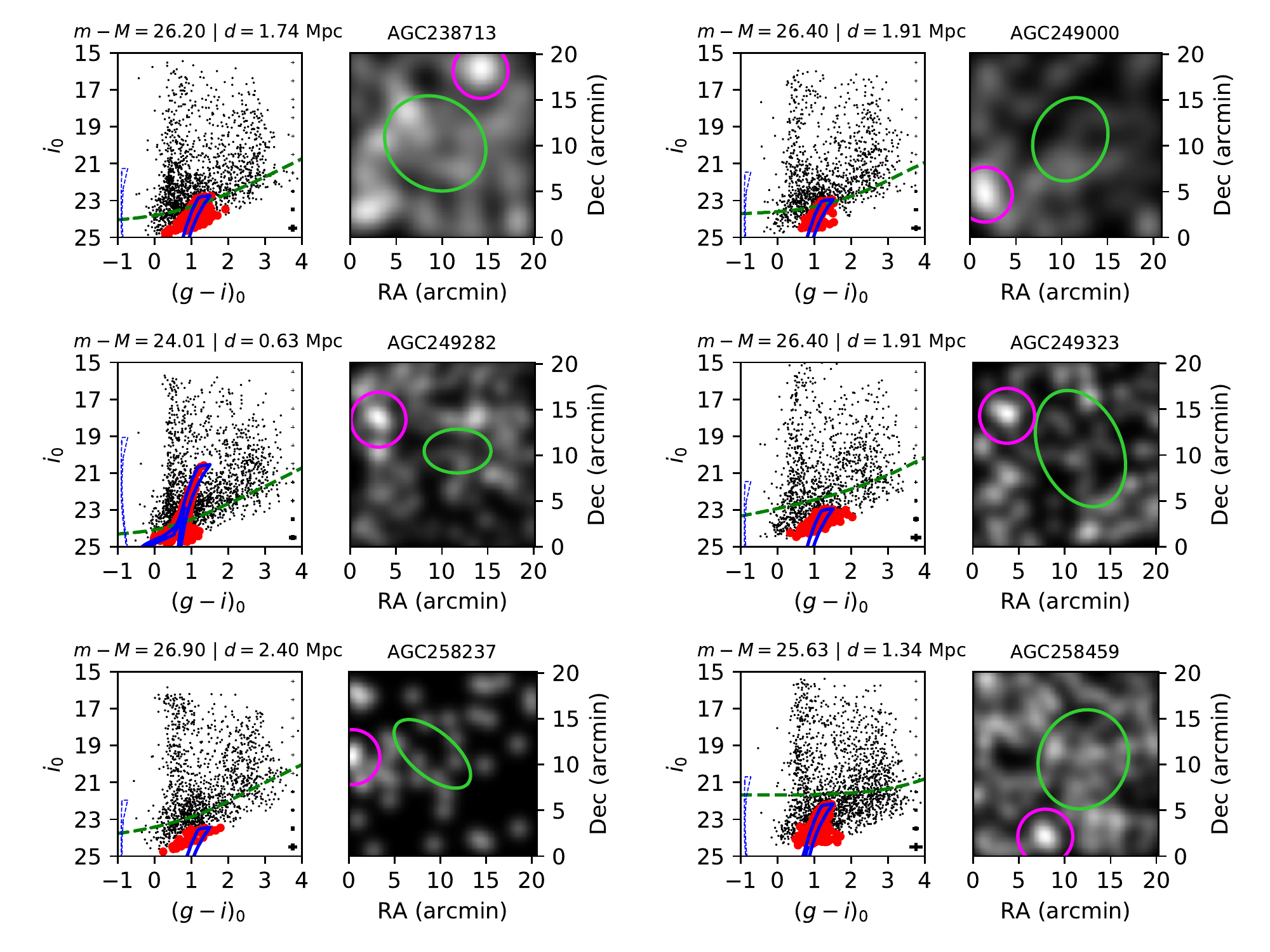}
  \caption{CMDs and smoothed stellar density maps, as described in Figure \ref{fig:results1}, for the remaining six of the fourteen objects objects that only had overdensities detected at the corner or edge of the image, several arc minutes away from the location of the ALFALFA UCHVC. See discussion of these objects in Section \ref{sec:nondets}. \label{fig:nondets2} }
\end{figure*}

For fourteen of the UCHVCs we imaged with the pODI camera, the most significant overdensity in the field was located in the corner or on the edge of the image. These overdensities had significances ranging from $\xi= 96.3\%$ to $\xi = 99.9\%$. While these overdensities are statistically significant when compared to a random distribution of objects, we do not consider them to be detections of stellar counterparts to the UCHVCs because they are too distant (approximately 9\arcmin\ to 14\arcmin\ away from the \hi\ detection) to be reliably associated with the gas.  To be fair, these overdensities could represent stellar populations that did once ``belong'' to the UCHVCs, but the gas and stars have been separated by some process. This hypothesis would be impossible to confirm without detailed follow-up observations. It seems more likely that the overdensities we have detected are groupings of unresolved, redshifted background galaxies. We emphasize that in order to assess the significance of our detections, we compare them to random spatial distributions, and objects in the universe are of course not spread over the sky randomly but instead tend to be clustered. Thus the Corner/Edge detections represent real overdensities of objects, but their distance from the \hi\ centroid makes it unlikely that they are actually associated with the UCVHCs. These fourteen objects are listed in the last grouping in Table \ref{tab:optical2} and, for completeness, we have included the CMDs and smoothed density maps for these objects in Figures \ref{fig:nondets1} and \ref{fig:nondets2}, but we do not carry out further analysis of these specific sources in this work.

One of the objects that is listed as a Corner / Edge detection only is AGC~226067.  This object was classified as a UCHVC in ALFALFA and we observed it with WIYN pODI, but we did not find a high-significance overdensity of point sources associated with the \hi\ gas distribution in the pODI images. Instead we detect an overdensity of point sources in the corner of the image only, $\sim 9\arcmin$ from the \hi\ centroid. There is, however, a diffuse blue object without any resolved stars near the center of the pODI images that we interpret as the optical counterpart to this UCHVC. In \citet{adams15a}, our group presented \hi\ synthesis mapping of AGC~226067 and showed that the area around this object includes a system with multiple components, some with \hi\ only and others with both \hi\ and optical emission. It was argued in that paper that the diffuse blue object that is an optical counterpart to AGC~226067 is at a distance comparable to the Virgo Cluster, in part because the stars in the object were unresolved in the pODI images. As described in the Introduction, the optical emission from AGC~226067 has been studied in detail by other groups and was found to likely be a pre-enriched remnant gas cloud stripped from another Virgo Cluster galaxy \citep{beccari17,sand17}.

\section{Discussion}
\label{sec:discussion}

\subsection{Comparison to Known Galaxies}
The optical counterparts listed in Table \ref{tab:optical} have absolute magnitudes that span the range of dwarf galaxies in the local Universe. Segue 1 \citep[$M_V = -1.5$;][]{belokurov07}, an ultra-faint dwarf galaxy, is at the low end of the range, with a similar magnitude to the faint limit for one candidate galaxy in our sample, AGC~215417. The range defined by the faint and bright magnitude limits for our candidate dwarf galaxies is $-1.4 > M_V > -7.1$.  A few dozen other dwarf galaxies, including some of the long-known ``classical'' dwarf spheroidal satellites in the Local Group, fall within this range \citep{mcconnachie12}. The eight candidate dwarf galaxies discovered in early Dark Energy Survey (DES) data \citep{bechtol15} also have magnitude estimates consistent with the range of our dwarf galaxy candidates, though nearly all of the classical satellites and UFDs discovered by wide-field surveys like SDSS and DES are within 350 kpc of the Milky Way.

A more relevant comparison may be to known gas-rich low-mass dwarf galaxies. Leo~T \citep{irwin07} has gas properties similar to those of the UCHVCs and an absolute magnitude of $M_V = -8.0$ \citep{dejong08}, and Leo~P is a UCHVC itself with $M_V = -9.3$ \citep{mcquinn15}. The M31 satellite LGS~3 has an \hi\ mass consistent with the UCHVCs ($3.8 \times 10^5$ \msun), but is brighter than any candidate galaxy presented in this paper at $M_V = -10.1$ \citep{mcconnachie12}. Pisces~A \citep{tollerud15} is also a UCHVC analog, with an absolute magnitude $M_V = -11.6$, a stellar mass $\sim 10^7$ \msun, and an \hi\ mass $\sim9 \times 10^6$ \msun\ \citep{tollerud16}. Pisces~A was cataloged by ALFALFA as AGC~103722 \citep{haynes18} and it is part of the final sample \citetext{J. Cannon, private communication, 2018} of the Survey of \hi\ in Extremely Low-mass Dwarfs \citep[SHIELD;][]{cannon11}, an ALFALFA follow-up program to explore gas-rich dwarf galaxies with \hi\ masses between $10^6$ and $10^7$ \msun and distances between 3 and 8 Mpc. Crucially, all four of these galaxies have evidence of recent or ongoing star formation, unlike the UCHVCs presented in this paper. A number of other galaxies with significant gas content, but without star formation, exist in the local Universe -- Antlia, Aquarius, KKR~3, KKH~86, and Phoenix all have at least $10^5$ \msun\ of \hi\ gas, but with absolute magnitudes between $-9.5 > M_V > -10.6$, all are brighter than the galaxy candidates described here \citep{mcconnachie12}. We also note that the \hi\ distribution in Phoenix is offset from the stellar distribution by about 6\arcmin, though the two distributions still overlap \citep{young07}.

A wide range of stellar masses has been observed in extremely faint galaxies. Several of the nearby UFDs discovered in SDSS, and all of the DES galaxy candidates, have stellar masses between $10^2$ and $10^3$ \msun, while the more gas-rich galaxies described above have stellar masses between $10^5$ and $10^6$ \msun. Though the observed mass function of gas-rich galaxies does not reach as low in mass as the UFDs, it does not preclude the existence of lower stellar mass gas-rich galaxies in the Local Group. UFDs within the virial radius of the Milky Way are extremely unlikely to have retained any of their gas due to ram pressure stripping \citep{spekkens14}, but more distant, gas-rich galaxies are less likely to have undergone such interactions. Since the galaxy candidates presented in this paper are located between 350 kpc and 2.2 Mpc, and are relatively isolated, it seems likely that they represent at least a portion of such a population of galaxies. 
 
\begin{figure*}
  \includegraphics[width=\textwidth]{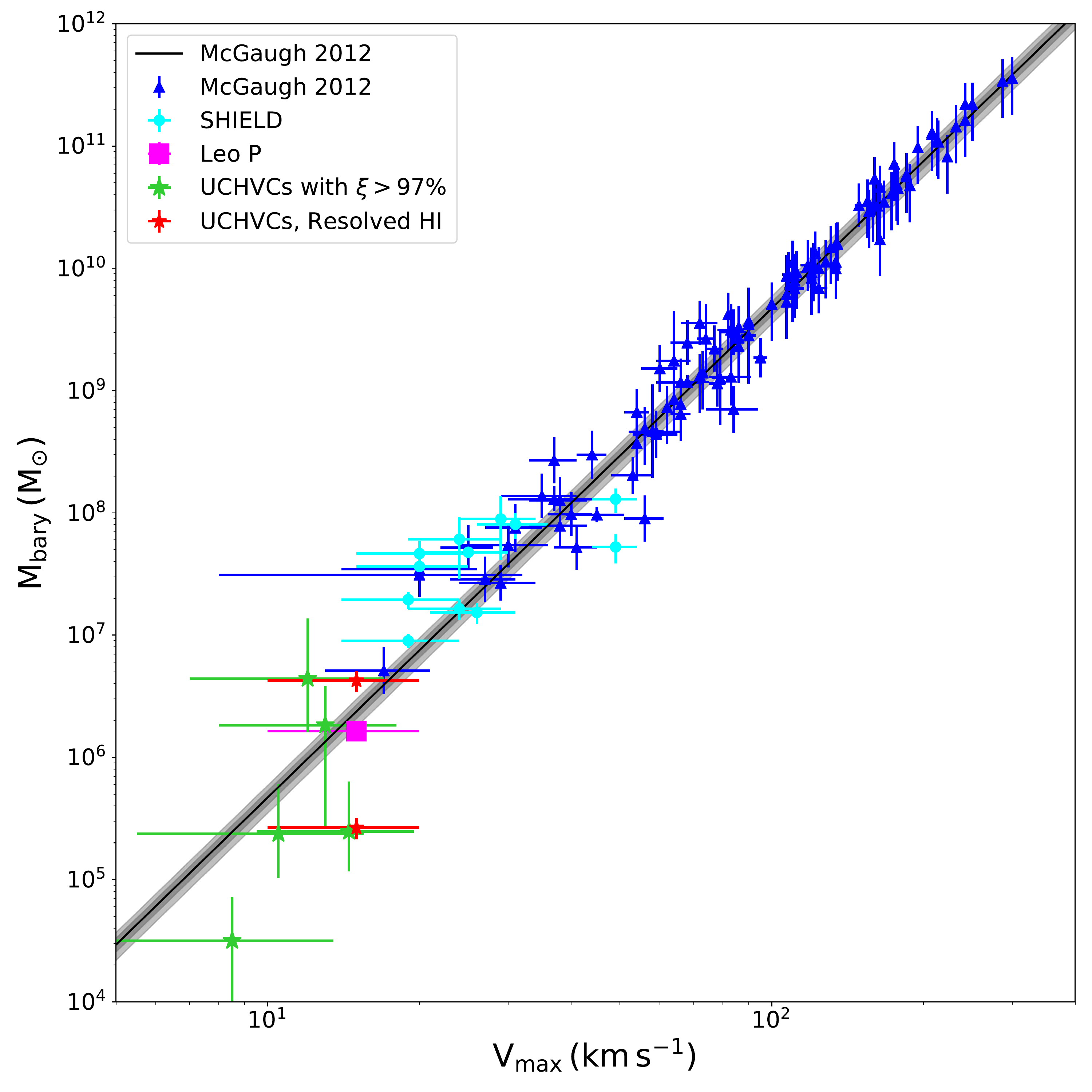}
  \caption{The baryonic Tully-Fisher relation, plotting the baryonic mass $M_{bary}$ against the maximum rotational velocity $V_{max}$, for galaxies from \citet[]{mcgaugh12} (blue) and the SHIELD survey \citep[][cyan]{cannon11,mcnichols16}, Leo~P \citep[][magenta]{bc14, mcquinn15}, and the UCHVCs with optical counterparts in the current paper. The UCHVC \hi\ properties are from VLA or WSRT (when possible, indicated with red) or ALFALFA (green). The UCHVCs are generally consistent with the BTFR defined by the higher-mass samples (black and grey shaded lines). \label{fig:btf} }
\end{figure*}

We can also compare the kinematic properties of the UCHVC galaxy candidates to other known galaxies of similar mass. Figure \ref{fig:btf} shows the UCHVCs with tentative optical counterparts in the context of the baryonic Tully-Fisher relation \citep[BTFR;][]{tf77, mcgaugh12}, an observed tight correlation between the baryonic mass of a galaxy and its maximum rotation velocity. In order to place the UCHVCs on this relation, we use the $w_{50}$ value from ALFALFA divided by a factor of two as a proxy for maximum rotational velocity. Given the uncertainty in the total stellar content of these galaxies depending on how the stellar luminosity is calculated (see Section \ref{sec:methods}), and the fact that the galaxies are nearly always gas-dominated, we use the sum of the total neutral gas mass (the \hi\ mass multiplied by a factor of 1.33 to correct for Helium abundance) and the mean stellar mass to represent the total baryonic mass, while the range of stellar mass and distance error both contribute to the error on the total mass.

The two UCHVCs shown in red in Figure \ref{fig:btf}, AGC~198606 and AGC~249525, have resolved \hi\ data from WSRT, including position-velocity slices along a clear velocity gradient that were used to constrain the maximum rotational velocity \citep{adams16}. 
In Figure \ref{fig:btf} we show both the BTFR and the gas-dominated galaxy sample from \citet{mcgaugh12}, and we add SHIELD galaxies \citep{mcnichols16} and Leo~P \citep{bc14} to populate the region between the McGaugh galaxies and the UCHVCs. The UCHVCs are broadly consistent with the BTFR, indicating that this relation may continue to the lowest baryonic masses and the extreme end of the distribution of galaxies that form in the Universe. However, below a rotational velocity of $\sim 10-15$ \kms, galaxies are no longer rotationally supported \citep{ds10, mcnichols16}, making the placement of the UCVHCs on the BTFR potentially inappropriate. The nature of the UCHVCs is still poorly understood, so additional \hi\ synthesis imaging will be crucial in order to investigate the gas distribution and dynamics and thus the interactions between the gas and stars, as well as to better determine the mass of the underlying dark matter halo. 

\subsection{Incidence Rate of Stellar Counterparts}
Of the 23 UCHVCs we observed with WIYN pODI, five ($\sim 20\%$) were found to have a stellar counterpart that appears to be associated with the \hi\ source (97.9\% to 99.9\% significance, and with a location close to the \hi\ centroid and/or within the \hi\ distribution). Four more UCHVCs had stellar populations detected at slightly lower confidence (92-95\%) but also in proximity to the \hi. Without higher resolution \hi\ synthesis imaging of all the UCHVCs, we cannot localize high peak column density regions that would be more likely to form stars. Therefore, we cannot make specific predictions about which UCHVCs should contain stellar populations.

Using hydrodynamic simulations of galaxy formation, \citet{sawala16} predict that fewer than 10\% of all halos with dynamical mass $<10^{8.5}$ \msun\ will have a luminous component. We can therefore make a general prediction about the number of UCHVCs that contain a stellar population, since all of the UCHVCs appear to have a dynamical mass $< 10^{8.5}$ \msun. Using Figure \ref{fig:sample} as a guide, we count the number of objects in the optical follow-up sample as a function of their mean column density: there are 9 UCHVCs with $\bar{N}_{HI} > 10^{19}$ atoms cm$^{-2}$ and 14 with $\bar{N}_{HI} < 10^{19}$ atoms cm$^{-2}$. If 10\% of these objects have a luminous component, following the prediction of \citet{sawala16}, we could expect 1 out of the 9 UCHVCs with $\bar{N}_{HI} > 10^{19}$ atoms cm$^{-2}$, and 1 to 2 out of the 14 UCHVCs with $\bar{N}_{HI} < 10^{19}$ atoms cm$^{-2}$, to have an associated stellar counterpart. Our current study indicates that the presence of a stellar counterpart in low mass halos is much more common, at least for gas-bearing halos.  We find that three of the nine UCHVCs with $\bar{N}_{HI} > 10^{19}$ atoms cm$^{-2}$ have stellar populations detectable in our images, and 5 out of 23 overall, substantially above the 10\% fraction of halos with luminous components predicted in the \citet{sawala16} models. Since the \citet{sawala16} model predictions are for \textit{all} halos, and not just halos containing a gaseous component, it might be that the neutral gas in these halos, and/or their isolation, helps explain the higher incidence rate.

Further, while the two highest column density sources in the pODI sample, AGC~198606 and AGC~249525, both have a $>97\%$ significance detection of an associated stellar population, the lowest column density source, AGC~215417, also has a $\sim 97\%$ detection. As a group, the high-significance detections evenly span the range of mean column density. However, localized regions of higher column density can exist in the gas distribution, even if the mean column density is measured to be below observed thresholds for star formation \citep[e.g.,][]{vanzee97}. Indeed, Leo~P, which has current star formation, has an ALFALFA-measured mean column density similar to that of the highest column densities in our sample ($\bar{N}_{HI} \sim 3 \times 10^{19}$ atoms cm$^{-2}$), though when observed with VLA, an order of magnitude higher peak column density was observed \citep[$N_{HI} \sim 2-6 \times 10^{20}$ atoms cm$^{-2}$;][]{leop1, bc14}. Leo~T, with recent but not active star formation, has a peak column density twice that of Leo P, $N_{HI} = 4.6 \times 10^{20}$ atoms cm$^{-2}$ \citep{adams17}. Though star formation may be unexpected in these isolated, low gas density systems, it clearly does occur in the local Universe, at least in some cases.

\subsection{Implications for Galaxy Formation Scenarios}
A growing population of extremely faint, gas rich galaxies in the Local Group and its environs presents an opportunity to test recent predictions about the formation of such systems. Because their optical properties are similar to those of the UFDs, one might speculate that the stellar components of the UCHVCs that we have identified could represent a progenitor class of galaxies to the UFDs. This is not a unique idea; \citet{willman06} and \citet{martin07} both speculate that the UFDs could be the end result of a tidal interaction between a dwarf galaxy and the Milky Way. \citet{br11} used $N$-body and semi-analytic models to show that the population of isolated UFDs in the Local Volume should be larger than previously observed, though they predicted most of these objects would be gas-poor as a result of stellar feedback effects or stripping. UCHVCs containing stars are even within the realm of simulation:  \citet{onorbe15} used high-resolution hydrodynamic simulations to predict the existence of objects with properties similar to the UCHVCs presented in this paper, with stellar masses between $10^4$ and $10^6$ \msun, and high gas-to-stellar mass ratios. \citet{verbeke15} and \citet{vandenbroucke16} inserted feedback from $60-300$ \msun\ Population III stars into $N$-body/hydrodynamical simulations, by injecting more energy from supernovae into the simulated interstellar medium while holding the metal yields for the supernovae constant. Those simulations produced several galaxies similar to the candidates presented here -- gas-dominated galaxies with stellar masses $\sim 10^5$ \msun, \hi\ masses $\sim 10^6$ \msun, and absolute magnitudes $M_V \gtrsim -9$. 

Clearly, further simulations and observations are needed to fully understand the UCHVCs as a population of faint dwarf galaxies. With no evidence of recent star formation in any of the candidate galaxies in this paper, a convincing explanation for turning off the star formation mechanisms while still retaining a significant reservoir of gas is necessary. The models outlined in \citet{verbeke15} and \citet{vandenbroucke16} relating to Population III stars have come the closest, though as those authors point out, the astrophysics of Population III stars are still poorly understood due to lack of observational evidence. 

In addition to further investigation of the origin and evolution of these objects, observational confirmation of the stellar populations associated with the UCHVCs is clearly needed. Observations with a larger-aperture ground-based telescope may be helpful for producing a deeper CMD and further investigating the veracity of the potential stellar counterparts we have identified. Spectroscopic observations of the stars in the overdensities, while only feasible with current capabilities for stars brighter than $i \sim 22$, would also be useful for confirming the association between the \hi\ gas and stars in the UCHVCs. Ideally, we would obtain space-based observations, similar to the deep HST imaging of Leo~P \citep{mcquinn15}, in order to explore the stellar content and star formation histories of these dwarf galaxy candidates. Finally, many more UCHVCs have already been observed with WIYN/ODI. Based on the incidence rate of candidates in this paper, we would expect to detect roughly two to six additional candidates in the ODI imaging data we have already acquired.


\acknowledgments
We thank the anonymous referee for their comments and suggestions, which improved the final manuscript.
We are grateful to the staff at the WIYN Observatory, Kitt Peak, and NOAO for their help during our various WIYN pODI observing runs. 
We also thank the ODI-PPA support scientists and programmers at WIYN, IU, and NOAO for their support and assistance with the reduction and analysis of the WIYN pODI data. 
The authors acknowledge the work of the entire ALFALFA collaboration for their efforts in support of the survey. 
W.F.J. and K.L.R. acknowledge support from NSF grant AST-1615483.
S.J. acknowledges support from the Australian Research Council’s Discovery Project funding scheme (DP150101734).
E.A.K.A. is supported by the WISE research programme, which is financed by the Netherlands Organisation for Scientific Research (NWO).
The ALFALFA team at Cornell is supported by NSF grants AST-0607007 and AST-1107390 to R.G. and M.P.H., AST-1714828 to M.P.H., and by grants from the Brinson Foundation. 
J.M.C. acknowledges prior support from NSF grant 1211683.
This research has made use of the NASA/IPAC Extragalactic Database (NED) which is operated by the Jet Propulsion Laboratory, California Institute of Technology, under contract with the National Aeronautics and Space Administration.

\facility{WIYN (pODI)}
\software{IRAF, PyRAF, Astropy, Numpy, Scipy, Matplotlib}

\clearpage

\end{document}